\documentclass[aps,amsmath,amssymb,superscriptaddress,showpacs,12pt]{revtex4}

\usepackage{graphicx}
\usepackage{dcolumn}
\usepackage{bm}
\usepackage{colordvi}
\usepackage{color}
\usepackage{ulem}

\begin{document}

\title{Regular and irregular dynamics of spin polarized wavepackets in a mesoscopic quantum dot at the edge of topological insulator}

\author{D.V. Khomitsky}
\email{khomitsky@phys.unn.ru}

\author{A.A. Chubanov}

\author{A.A. Konakov}

\affiliation{Department of Physics, National Research Lobachevsky State University of Nizhni Novgorod, 603950 Gagarin Avenue 23,
Nizhni Novgorod, Russian Federation}

\begin{abstract}

The dynamics of Dirac-Weyl spin-polarized wavepackets driven by periodic electric field is considered for the electrons in a mesoscopic quantum dot
formed at the edge of two-dimensional HgTe/CdTe topological insulator
with Dirac-Weyl massless energy spectra, where the motion of carriers is less sensitive to disorder and impurity potentials. It was observed that the interplay of strongly coupled spin and charge degrees of freedom creates the regimes of irregular dynamics both in coordinate and spin channels. 
The border between the regular and irregular regimes determined by the strength and frequency of the driving field is found analytically within the  quasiclassical approach by means of the Ince-Strutt diagram for Mathieu equation, and is supported by full quantum mechanical simulations of the driven dynamics.
The investigation of quasienergy spectrum by Floquet approach reveals the
presence of non-Poissonian level statistics which indicates the possibility of chaotic quantum dynamics and corresponds to the areas of parameters for irregular regimes within the quasiclassical approach.
We found that the influence of weak disorder leads to partial suppression of the dynamical chaos. Our findings are of interest both for progress in a fundamental field of quantum chaotic dynamics and for further experimental and technological applications of spin-dependent phenomena in nanostructures based on topological insulators.

\pacs{05.45.Mt, 75.70.Tj, 85.75.-d}

\end{abstract}

\maketitle

\section{Introduction}

Recently a growing attention in various field of fundamental science  has been started to focus on the properties of so called Dirac-Weyl fermions \cite{VolovikBook}. We can mention here just few examples:
in elementary particle physics massless neutrinos are described as Weyl fermions \cite{Pal}, in quantum optics laser-induced excitations in the system of ultracold atoms in optical superlattices have such properties \cite{Lan} and, of course, numerous manifestations in condensed matter physics \cite{Vafek}.
Interestingly, that in solid state systems Dirac-Weyl fermions can have different dimensionality, such as 3D low-energy long-wavelength
 excitations in Weyl semimetals \cite{Wan}, well-known 2D electron and hole excitations near Brillouin zone $K$ point in  graphene \cite{Beenakker} as well as on the surface of 3D topological insulators (TI) \cite{TI} and 1D Weyl fermions at the edge of  graphene \cite{Volkov2009} or at the edge of 2D TI \cite{TI}.

In the present work we are interested in properties of simplest 1D Dirac-Weyl fermions, which for a case of free particles can be described by the Hamiltonian of the form

\begin{equation}
H_0=\hbar v_F k_i \sigma_j,
\label{hweyl}
\end{equation}

where the parameter $v_F$ is the electron velocity at the Fermi level, and $k_i$ and $\sigma_j$ are components (in general different) of
the wavevector and Pauli spin vector, respectively, both depending on the system geometry. Pauli vector defines that the wave function is two-component,
but its physical meaning depends on the realization of Weyl fermions, and it could represent spin, pseudospin or, generally speaking, some other
"effective spin" degree of freedom. Despite the simplicity of the Hamiltonian (\ref{hweyl}), it maintains all general properties of the Dirac-Weyl
fermions, namely, its linear dispersion and "two-bandness". The 1D excitations described by the Hamiltonian (\ref{hweyl}) can be most easily
produced at the edge of 2D topological insulator, so in the following we will assume this special case for definiteness.
These materials represent the condensed matter state with bulk band gap and propagating edge or surface states which are protected
from backscattering by time-reversal symmetry and have energies within the bulk gap. Thus, in 2D TI an efficient transport through 1D edge
channels can be produced.

While static properties of Dirac-Weyl fermions at the edge of 2D TIs are well described \cite{TI}, much less is known about evolution of the Dirac states
under external driving fields. The dynamics of a quantum system often can be classified as one of two limiting cases: the few level dynamics, and the
evolution involving very many levels, usually referred as quasiclassical dynamics. In the latter case there are commonly used analogies between the quantum evolution and the corresponding dynamics of a classical counterpart. The problem for application of this approach to Dirac-Weyl fermions is that there is no direct classical counterpart available. The same questions arise during the investigation of special regimes of dynamics called irregular, stochastic or chaotic dynamics.
It is still an open question which type of dynamics can be called as quantum
chaotic dynamics when a systems lacks a direct classical counterpart.
Several approaches to quantum stochasticity have been successfully developed
for many years \cite{Gutzwiller,Reichl,Stockmann}. First of all, in a quantum system with the high number of energy level $N \gg 1$ the quasiclassial approach can be applied, regardless on whether the system has a classical counterpart.
The general rule states that a classically chaotic systems will demonstrate certain chaotic dynamics also in the quantum regime. However, it is not clear whether any quantum system without the classical counterpart such as Dirac-Weyl fermion will demonstrate stochastic behavior under particular circumstances. Several approaches aimed on establishing connections between quantum and classical systems with chaos have been proposed,
including the studies of irregular dynamics in condensed matter systems \cite{Stockmann}, and quantum dots \cite{Nakamura}. The connections between classical and quantum chaotic systems have been established in such properties like the structure of the quasienergy spectra and the phenomena of quantum diffusion in the Hilbert space.
Here the analogies between the diffusion along the resonance eigenstates
and along the separatrices in corresponding classical system have been found \cite{Reichl,Stockmann}, including the analogue of Arnol'd diffusion
in quantum systems subject to periodic driving \cite{DIM}.

Among the simplest and important systems without the classical counterpart are  low-dimensional structures where the spin is strongly coupled to the orbital degrees of freedom, including those with spin-orbit coupling (SOC) and surfaces of TIs. The importance of semiconductor structures with strong SOC has been recognized during the last decade, and a significant progress can be observed
in corresponding field of nanophysics called spintronics \cite{Zutic,Dyakonov}. We may expect similar rich variety of the results for the condensed matter systems with Dirac cones in the electron spectra as the field of topological insulators continues to grow \cite{TI}.

The presence of SOC leads to correlation between space and spin degrees of freedom and, thus, creates a possibility of non-trivial dynamics or evolution
in coupled coordinate and spin channels.
It is known that, for example, the combined effects of SOC and
the resonance in a multi-level system subject to strong driving may lead to unusual nonlinear behavior in well-known regimes such as quasiclassical dynamics of the electron in a double quantum dot \cite{KS2009}, or the Rabi frequency dependence on the driving strength in the electric dipole spin resonance in a double quantum dot \cite{Rabi2012}.
It was shown also that in double quantum dot with SOC other interesting regimes can develop such as phase synchronization,
or even chaotic spin-dependent dynamics \cite{Chotorlishvili}.
Other examples can be found in a 2D mesoscopic semiconductor quantum dot with SOC \cite{Berggren}, and in the 2D deformed harmonic oscillator potential with SOC \cite{Zinner2014}.
In these studies the non-Poissonian level statistics has been found which indicates the presence of quantum chaos.
In our recent paper \cite{billiard2013} we have found the development of strongly irregular dynamics in this system under the periodic driving by the electric field, which manifested itself in both charge and spin channels.
Such nontrivial spin-dependent evolution of quantum states should also develop in the driving dynamics of electrons on the surface of 3D TIs
and at the edge of 2D TIs, where the Dirac-Weyl Hamiltonian can be described as the limiting case of the electron Hamiltonian with extremely high
linear-in-k SOC.

The major problem of establishing the quantum-classical correspondence in such spin-dependent systems is the mentioned absence of a direct classical
counterpart which creates obstacles on describing such systems in terms of the classical chaotic dynamics. So, one needs to find some techniques
that allows to distinguish between regular and irregular dynamics in purely quantum system. The primary tools for overcoming such
difficulties are the Floquet analysis for periodic driving  \cite{Reichl,Stockmann,DIM}
and the analysis of transport properties reflecting the regular or chaotic structure of energy spectrum and eigenstates \cite{Nakamura,BSSR}.

In the Floquet analysis one may look on the degree of delocalization of the Floquet eigenstates in the Hilbert space
of the basis states of the bounded quantum system
or the quasienergy level statistics clearly indicating the possibility of diffusion and chaos development \cite{Reichl,Stockmann},
and on the direct Fourier analysis of the observables, or quantum mean values \cite{Gutzwiller,Yang2007}.
Other tools include the analysis of Poincare sections built in various pairs of coordinates for both coordinate and spin degrees of freedom,
not necessarily the canonically conjugated ones \cite{billiard2013,Yang2007}, or the tracking of the evolution for the variance for the number of energy
levels involved in the dynamics. Here the growth of such variance indicates the development of chaotic regime, and the saturation points
on the transition to the quasi-regular mode with finite number of levels participating in the evolution \cite{DIM,billiard2013}.

In the present paper we address the complex driven dynamics of Dirac-Weyl wavepackets representing the electrons localized in
mesoscopic structures formed at the edge of HgTe/CdTe 2D topological insulator by magnetic barriers. Such barriers are required due
to the effect of Klein tunneling prohibiting the purely electrostatic confinement of the Dirac-Weyl fermions with the Hamiltonian (\ref{hweyl}).
Our general aim is to find out whether the dynamics of a Dirac-Weyl wavepacket with Hamiltonian (\ref{hweyl}) in a quantum dot (QD) formed at the edge
of TI is regular or irregular, if the packet is driven by a monochromatic electric field. To find this, we consider the time evolution of various
observables associated with the wavepacket dynamics, their Fourier spectra, and the "phase space" portraits of different pairs of variables,
both for coordinate and spin degrees of freedom.
We find that certain properties of driven evolution are sustained for wavepackets of different shape and are not smeared
by a moderate disorder potential.

This paper is organized as following.
In Section II we introduce a model of quantum states in 1D QD at the edge of 2D TI based on HgTe/CdTe quantum well for the case of magnetic barriers
with finite transparency where the wavefunctions have nonvanishing tails inside the barriers. We consider a case of a macroscopic QD
with a length $L=3$ mkm in order to obtain a large number of levels ($N_{\rm{max}} \approx 100$) in the TI bulk gap
which is desirable to capture the quasiclassical traits of chaos development. Such assumption of a long 1D mesoscopic QD is feasible
since the experiments report rather high values of mean free paths in such structures, reaching several microns \cite{TI}.
In Section III we perform a quasiclassical analysis of the driven dynamics and find the correspondence between the equations for the spin dynamics and the Mathieu equation, and identify the associated instability borders indicating the possible onset of chaotic dynamics.
In Section IV we describe the Floquet eigenstates which may indicate the diffusion in the Hilbert space showing the possibility of chaotic dynamics.
In Section V we consider the evolution in the clean limit (no static disorder or noise) for the electron inside the QD, where the electron is represented via the spin polarized wavepacket. 
In this Section we consider the initial wavepacket which is described by a wide envelope function in the coordinate space and by a narrow distribution in the Hilbert space of eigenstates of the unperturbed Hamiltonian.
Such narrow distribution allows drawing the correspondence between the full quantum mechanical treatment from this Section and the quasiclasical approach described in Section III.
We consider the evolution under monochromatic driving electric field, and describe it in terms of phase space plots generalized also for pairs of non-conjugate spin variables, Fourier spectra, diffusion in Hilbert and coordinate space, and Lyapunov exponent. 
In Section VI we add the random disorder potential representing the non-ideal character of real nanostructure
as well as possible noise in the system and study the driven evolution here. 
In this Section we take a narrow initial wavepacket which has a wide distribution along the basis states of the unperturbed Hamiltonian, making valid only the full quantum mechanical treatment. We find that the qualitative features of the driven evolution are the same for both types of wavepackets considered here and in the previous section, which provides additional justification for the quasiclassical treatment in Section III.
Finally, in Section VII we present our conclusions.

\section{Model for quantum states}

One of the first examples of Dirac-Weyl fermions in condensed matter were the edge states in the HgTe/CdTe quantum wells where the tuning of
the well width may create the phase where topologically protected edge states exist \cite{TI,BHZ}.
It is known, however, that the applications of TI in nanoelectronic devices require the fabrication of localized small-to-medium size object like quantum
dots. Several models of QD formation at the edge of TI have been proposed during the last years \cite{Timm,QDTI}. Most of them relevant to 1D QD at the edge of 2D TI deal with simplified assumptions of non-transparent magnetic
barriers which are required to confine the electrons with massless Dirac-Weyl spectrum \cite{TI}.
It should be mentioned that the similar methods of confinement by creating a gap in the spectrum by  magnetic field or other mass terms
have been offered also in other materials with Dirac-Weyl spectra such as graphene where offered \cite{grapheneqd}.
Under such assumptions the spectrum of discrete energy levels inside QD forms a set of equidistant levels located in two ladders above and below
the Dirac point of TI where two linear dispersion branches cross \cite{Timm}. For each level the corresponding eigenstate is a two-component spinor
with certain spin polarization, which makes this system a promising candidate for studying there a driven dynamics excited by external
electric field tuned to match the interlevel resonance splitting.

In our paper we use the envelope function approximation with the effective Hamiltonian for the 1D electron in a QD confining
the propagating spin-polarized states at the edge of 2D HgTe/CdTe TI:

\begin{eqnarray}
H_{QD}=\hbar v_F k_y \sigma_z - M_0 \theta(-y) \left(\sigma_x \cos \theta_0 + \sigma_y \sin \theta_0 \right) - \\
- M_L \theta(y-L) \left(\sigma_x \cos \theta_L + \sigma_y \sin \theta_L \right).
\label{hqd}
\end{eqnarray}

Here the first term is the effective Dirac-Weyl Hamiltonian (\ref{hweyl})
for unperturbed gapless edge states on the boundaries of the TI.
The Fermi velocity $v_F$ is determined by the HgTe layer thickness, and for our model we take the value $v_F=5.3 \cdot 10^7$ cm/s and consider the band gap in HgTe/CdTe to be around $40$ meV in the inverted regime which corresponds to the quantum well width in the range of $8 \ldots 9$ nm \cite{TI}.
The second and third terms in (\ref{hqd}) describe local exchange interaction between electron near edge of the quantum well and the magnetization of the magnetic stripes. The magnetization of both contacts
are assumed to be uniform without any domain structure. This situation is typical for nanomagnets with size less than 100 nm at least in one direction \cite{MMbooks}.
The barrier magnitudes $M_0$ and $M_L$ can be viewed as exchange energies. Both magnetic contacts are located along the TI edge at $y=0$ and $y=L$
forming a 1D QD with the width $L$, as it is shown schematically in Fig.\ref{figmodel}(a).
The QD length $L=3$ mkm in our model is sufficiently large to justify the application of the envelope function approximation.
We consider the barriers with finite transparency by choosing finite amplitudes $M_0$ and $M_L$ which are taken as to cover the whole band gap of the HgTe/CdTe quantum well.
The size of the magnets along the TI edge is considered to be comparable with the QD size, so we can assume them to be infinite in this direction
since the wavefunction of the QD states decays exponentially into barriers on the size which is substantially smaller than the magnet length,
as we shall see below. The angles $\theta_0$ and $\theta_L$ describe the orientation of the magnetization in the left and right barriers, respectively. Our Hamiltonian is a generalization of the previously derived model for QD with impenetrable barriers \cite{Timm} for the more realistic case of
the barriers with finite transparency reflected in their finite height $M_{0,L}$.
It should be noted that usually the few-electron regimes are desirable for operations of QD as qubit or other information
processing device. So, we believe that it is necessary to use dielectric magnetic materials in order to prevent the excessive leak of the electrons
from the leads into the QD.

The stationary 1D Schr\"odinger equation $H_{QD}\Psi=E\Psi$ for the two-component envelope function $\Psi=(\psi_1(y),\psi_2(y))$
is accompanied by the boundary conditions at $y=0$ and $y=L$ which can be derived from its integration over infinitesimal small
region near the boundary, yielding the requirements

\begin{eqnarray}
\Psi(-0)=\Psi(+0), \\
\Psi(L-0)=\Psi(L+0)
\label{bc}
\end{eqnarray}

meaning that the envelope function has to be continuous at the boundaries between the QD and the barriers. The spatial dependence
of the solution for a confined state with energy $E<(M_0,M_L)$ inside the barriers at $y<0$ and $y>L$ has the form of decaying
under-the-barrier exponents,

\begin{equation}
\Psi_{y<0}=B
\left[
\begin{array}{c}
1
\\
-\frac{i \sqrt{M_0^2-E^2}+E}{M_0}e^{i \theta_0}
\end{array}
\right]
\exp \left(\frac{\sqrt{M_0^2-E^2}}{\hbar v_F}y \right),
\label{psileft}
\end{equation}

\begin{equation}
\Psi_{y>L}=D
\left[
\begin{array}{c}
1
\\
\frac{i \sqrt{M_L^2-E^2}-E}{M_L}e^{i \theta_L}
\end{array}
\right]
\exp \left(-\frac{\sqrt{M_L^2-E^2}}{\hbar v_F}y \right),
\label{psiright}
\end{equation}

and the eigenstate inside the QD is the spinor with real wavenumber in its exponents,

\begin{equation}
\Psi_{QD}=
\left[
\begin{array}{c}
C_1 e^{i E y/\hbar v_F}
\\
C_2 e^{-i E y/\hbar v_F}
\end{array}
\right],
\label{psidot}
\end{equation}

where the coefficients $B,C_1,C_2,D$ are determined from the boundary conditions (\ref{bc}), and the energy $E$ is found from
the corresponding secular equation with Hamiltonian (\ref{hqd}).
It should be mentioned that the states (\ref{psileft})- (\ref{psidot}) correspond to the spin polarization
which is always in the plane of the 2D TI, that is, $S_z=0$.

The system of equation (\ref{bc}) for the envelope function (\ref{psileft})- (\ref{psidot}) can be solved analytically for the case
of non-transparent barriers where the wavefunction does not enter the under-the-barrier region \cite{Timm,QDTI} where
a sequence of up- and down- strictly equidistant energy levels

\begin{equation}
E^{(0)}_{n_0}=\Delta E^{(0)} \left(n_0+\frac{1}{2}+\frac{\theta_L-\theta_0}{2\pi}\right),
\label{en0}
\end{equation}

($n_0= \pm 1, \pm 2, \ldots$) is formed with spacing which is independent of $\theta_L-\theta_0$,

\begin{equation}
\Delta E^{(0)} = \frac{\pi \hbar v_F}{L}.
\label{den0}
\end{equation}

In the present paper we will consider the case of parallel orientation $\theta_0=\theta_L=0$ since from (\ref{en0}),(\ref{den0}) it follows that
different angles of magnetization inside the barriers define mostly the internal structure of corresponding eigenstates and their spin polarization inside
the QD, and do not affect the level spacing for the idealized case of non-transparent barriers which determines in our model the primary frequency of driving field.
This result is preserved in the case of transparent magnetic barriers considered in our model (\ref{hqd}).
One may expect that various difference between $\theta_0$ and $\theta_L$ will lead to the formation of quantum states with different spatial symmetry, but for our dynamical problem this will produce only quantitative effects on the structure of matrix elements of the external perturbation, and thus only minor effects on the dynamical properties which are in the focus of our studies.
Besides, we choose equal amplitudes of magnetic barriers $M_0=M_L$ which creates
a QD with symmetric potential profile, although various combinations of $M_0$, $M_L$, $\theta_0$, and $\theta_L$ can be equally considered if other materials and/or experimental setups are chosen.

In our model the spectrum cannot be found analytically, and has to be obtained from a transcendental equation, which leads in general to a non-equidistant spectrum with non-uniform level spacing $\Delta E$.
However, for the mesoscopic QD with $L=3$ mkm where from (\ref{den0}) $\Delta E^{(0)} \approx 0.38$ meV and the condition $\Delta E^{(0)} \ll M_{0,L}$ is satisfied meaning that there are many levels below the barriers are present ($N_{\rm{max}} \approx 100$), we have the level spacing $\Delta E$ being very close to the equidistant value $\Delta E^{(0)}$ from (\ref{den0}).

The scheme of the discrete energy levels inside the QD is presented in Fig.\ref{figmodel}(b) with a large interlevel distance which is shown as a guide to the eye and not to scale. Together with the discrete levels we
plot the linear dispersion branches of the Weyl Hamiltonian (\ref{hweyl}) describing the edge states \cite{TI} before the confining barriers
are applied, together with corresponding $z$-aligned spin mean values $S_z$ and the boundaries of the bulk energy gap $E_g = 40$ meV.
This gap allows to limit the barrier width by $M_0=M_L=E_g/2$ since only the edge states within the bulk gap are relevant for the edge QD
where they are not masked by the bulk states.

Our final task considering the model of quantum states inside the QD is the choice of the localized initial condition for the dynamical problem
representing a spin-polarized electron which has been injected through one of the magnetic barriers into the dot. We model such condition by a Gaussian
wavepacket with two different widths and center locations with their probability density distribution shown in Fig.\ref{figmodel}(c)
by the solid line (1) and dashed line (2). In terms of the spatial size the packets widths are 1 and 0.1 microns, respectively, which are
reasonable values for the semiconductor structures being considered where the mean free path for the electron is about 3 microns \cite{TI}.
We refer to these packets as wide and narrow, respectively, throughout the following text, and consider the zero mean value of the initial quasimomentum $\overline{k_y}(0)=0$.
The spin polarization for the corresponding spinor representing the initial packet is chosen as to coincide with the magnetization of
the magnetic barrier (or electrode) from which the packet has been injected, that is, the $S_x=1$ polarization of the left barrier,
since the majority of the electrons traveling through the magnetic materials without special tuning usually gain the polarization from the host material.
The initial condition $\Psi_0 (y)$ needs to be decomposed over the basis states $\Psi_n(y)$ inside the QD for further treatment of its evolution, that is, the coefficients $C_n$ in the decomposition $\Psi_0=\sum_n C_n \Psi_n(y)$ have to be found by standard methods. The structure of their absolute value distribution $|C_n|^2$ in the space of basis states is shown in Fig.\ref{figmodel}(d) and (e) for two initial packets from Fig.\ref{figmodel}(c), respectively. As expected, a wider packet (1) in real space is described by a narrow distribution of $|C_n|^2$ in the Hilbert space compared to the narrow packet (2). We will consider the driven dynamics for both types of wavepackets (1) and (2) in the next sections, and we will see that the difference in their shape in coordinate
or Hilbert space leaves certain dynamical features qualitatively the same which allows to consider our finding as relevant for various types of initial conditions.

\section{Quasiclassical dynamics}

We start by the application of the quasiclassical approach to the driven dynamics for the Hamiltonian

\begin{equation}
H=H_{\rm{QD}}+V(y,t)
\label{hfull}
\end{equation}

where $H_{\rm{QD}}$ is the stationary Hamiltonian (\ref{hqd}), and $V(y,t)$ is the driving term describing the electric field inside the quantum dot at $0<y<L$,

\begin{equation}
V(y,t)=e\mathcal{E}_0 y \cos \omega_0 t,
\label{driving}
\end{equation}

where $e$ is the elementary charge, $\omega_0$ is the driving frequency, and $\mathcal{E}_0$ is the electric field strength. It corresponds to the spatially uniform and harmonic electric field directed along the TI edge. This field can be generated by additional electrostatic gates arranged close to the bounding magnetic barriers. We consider the quasistationary field which is produced by the modulation of the gate potential and is considered as being spatially uniform on the scale of the QD $L=3$ mkm.

The mesoscopic size of our QD (3 mkm) and the large number of energy levels (about 100) allows the application of the quasiclassical approach. Such approach can be applied both for systems with and without the direct classical counterpart, including the ones with spin-orbit coupling in nanostructures if their dimensions generate the quasiclassically high number of energy levels \cite{KS2009}.
In the framework of this approach we consider only the evolution of quantum mechanical mean values. The evolution of the mean value ${\overline x}(t)$ corresponding to the time-independent operator $x$ is governed by the equation

\begin{equation}
\frac{d \overline{x}(t)}{dt}=\frac{i}{\hbar} \overline {\left[H, x \right] }
\label{evol}
\end{equation}

where $\overline{(\ldots)}$ stands for quantum mechanical averaging in a given state $\Psi({\bf r}, t)$. 
By applying Eq.(\ref{evol}) to the dynamics of the average of the operator product $\overline{AB}$ one can treat it as a product of two averages $\overline{A} \cdot \overline{B}$ if the distribution of the coefficients $C_n$ for the wavefunction decomposition via the basis states $\phi_n$ is a narrow function centered at certain $\overline{n} \gg 1$. It means that the width $\delta n$ satisfies the condition $\delta n \ll \overline{n}$, although being quasiclassically high, $\delta n \gg 1$, as it is considered in quasiclassical approach applied in this section. Our full quantum mechanical simulations presented in the following sections support this approximation on the parameters of the evolution. In particular, in Section V we perform the full quantum mechanical simulation for the wavepacket which is initially described by a narrow distribution in the Hilbert space, and satisfies the criteria for the quasiclassical treatment mentioned above. By contrast, in Section VI we take the initially wide packet in the Hilbert space for which only the full quantum mechanical simulation is applicable. We will see  that in both cases the qualitative results for the evolution are largely the same, which is a strong evidence of  correspondence between the quasiclassical and full quantum approaches presented in our paper.

For the Hamiltonian (\ref{hfull}) with the parallel orientation of the magnetic barriers $\theta_0=\theta_L=0$ one obtains from (\ref{evol}) the following set of equations defining the evolution of the coordinate and spin mean values inside the quantum dot:

\begin{subequations}
\begin{eqnarray}
\frac{d \overline{y}(t)}{dt} =v_F \overline{\sigma_z}(t) \label{sysevol-a} \\
\frac{d \overline{k_y}(t)}{dt}=\frac{\omega_b}{2}\frac{\partial F_b}{\partial y}\overline{\sigma_x}(t)-\frac{e\mathcal{E}_0 }{\hbar} \cos \omega_0 t \label{sysevol-b} \\
\frac{d \overline{\sigma_x}(t)}{dt}=-2 v_F \overline{k_y}(t) \overline{\sigma_y}(t) \label{sysevol-c} \\
\frac{d \overline{\sigma_y}(t)}{dt}=2 v_F \overline{k_y}(t) \overline{\sigma_x}(t) + \omega_b F_b(y) \overline{\sigma_z} (t)  \label{sysevol-d} \\
\frac{d \overline{\sigma_z}(t)}{dt}=- \omega_b F_b(y) \overline{\sigma_y} (t)  \label{sysevol-e}
\end{eqnarray}
\end{subequations}

The frequency $\omega_b=2 M_0/\hbar$ and the function $F_b(y)=\Theta (-y)+\Theta (y-L)$ are associated with the presence of magnetic barriers. Although the function $F_b(y)$ is nonzero only in the barrier regions $y<0$ and $y>L$ while we study the quasiclassical dynamics within the quantum dot at $0<y<L$, one needs to keep it at least in the equation (\ref{sysevol-e}) since it determines the evolution of the $\sigma_z$ spin component.
This means that the coupling to the magnetic barriers is essential for the driven dynamics to be initiated, which is confirmed by the full quantum mechanical calculations in the next sections.

The system (\ref{sysevol-a}) - (\ref{sysevol-e}) is the system of differential equations with non-stationary coefficients, and cannot be solved analytically in general case. This is a typical situation: for example, in our paper \cite{KS2009} we have performed mostly a computational analysis for the similar problem. However, for some regimes of driving certain analytical results can be obtained. For example, the harmonic time dependence of the driving term (\ref{driving}) allows to integrate equation (\ref{sysevol-b}) directly
inside the QD region $0<y<L$ where $F_b \equiv 0$ yielding

\begin{equation}
\overline{k_y}(t)=\overline{k_{y0}}-\frac{e\mathcal{E}_0}{\hbar \omega_0}\sin \omega_0 t.
\label{kyt}
\end{equation}

Having defined the time dependence of $\overline{k_y}(t)$ one can see that the other equations (\ref{sysevol-a}), (\ref{sysevol-c}) - (\ref{sysevol-e}) are linear equations with periodic coefficients with respect to the time variable. It is known that such equations can demonstrate unstable solutions which is often referred as parametric resonance. Since all of the coordinate and spin variables are coupled through the equations in the system, it is enough to determine the border of such instability for at least one variable. If a variable demonstrates an irregular time dependence, the other coupled variables will also acquire similar behavior with time. The most simple analytical result leading to a well-known type of equation with instabilities can be made for the spin variable, namely, for $\overline{\sigma_x}(t)$. By differentiating the left and right sides of (\ref{sysevol-c}) we obtain the equation of the second order,

\begin{equation}
\frac{d^2 \overline{\sigma_x}}{dt^2}+2 v_F \overline{k_y} \frac{d \overline{\sigma_y}}{dt}+ 2 v_F \frac{d \overline{k_y}}{dt} \overline{\sigma_y}=0.
\label{secsigx}
\end{equation}

We put the right hand side of (\ref{sysevol-d}) for the time derivative of $\overline{\sigma_y}$ into (\ref{secsigx}). 
We neglect the term with $F_b(y)$ for the following quasiclassical treatment of the motion inside the dot since it is present only in the barrier regions.
Then we substitute into (\ref{secsigx}) the time dependence for $\overline{k_y}(t)$ from (\ref{kyt}), and by using (\ref{sysevol-d})
we replace $\overline{\sigma_y}$ by $\frac{d \overline{\sigma_x}(t)}{dt}/(-2v_F \overline{k_y(t)})$. As a result, we arrive to the equation for the $\overline{\sigma_x}$ alone:

\begin{equation}
\frac{d^2 \overline{\sigma_x}}{dt^2}+f(t)\frac{d \overline{\sigma_x}}{dt}
+g(t) \overline{\sigma_x}=0,
\label{eqsigmax}
\end{equation}

where

\begin{equation}
f(t)=-\frac{1}{\overline{k_y(t)}}\frac{d\overline{k_y(t)}}{dt},
\quad
g(t)=4v_F^2\overline{k_y(t)}^2.
\label{fg}
\end{equation}

The first derivative of $\overline{\sigma_x}$ in (\ref{eqsigmax}) can be eliminated by introducing the new variable $\sigma_1(t)$ according to

\begin{equation}
\overline{\sigma_x}=\sqrt{\overline{k_y(t)}} \sigma_1(t).
\label{sigma1}
\end{equation} 

It is essential to note that our analysis in this section holds is appropriate 
in the region $\overline{k_y(t)}>0$ which means by checking (\ref{kyt}) that the moderate electric field is allowed which does not lead to negative values of  
$\overline{k_y(t)}$ starting from the quasiclassically high initial value of $\overline{k_{y0}}$. Then we introduce the dimensionless initial momentum as

\begin{equation}
k_0=\frac{\overline{k_{y0}} \pi \hbar v_F}{e\mathcal{E}_0 L},
\label{k0}
\end{equation}

and we also define the electric field-dependent frequency

\begin{equation}
\Omega=\frac{2 v_F e \mathcal{E}_0}{\hbar \omega_0}.
\label{omelec}
\end{equation}

We note that due to the confition $\overline{k_y(t)}>0$ and the expression (\ref{kyt}) our analysis is restricted to the area where $k_0>1$ which, according to (\ref{k0}), means that one must stay within in the quasiclassical region of moderate electric field.
Taking into account (\ref{sigma1}), (\ref{k0}), (\ref{omelec}), and by introducing the dimensionless time variable $\tau$ as

\begin{equation}
\omega_0 t=\tau
\label{tau}
\end{equation}

we transform Eq.(\ref{eqsigmax}) to the second order differential equation for $\sigma_1(\tau)$ which has the following form:

\begin{equation}
\frac{d^2 \sigma_1}{d \tau^2}+ \Theta(\tau) \sigma_1=0,
\label{hill}
\end{equation}

where

\begin{equation}
\Theta(\tau)=\frac{\Omega^2}{\omega_0^2} \left(k_0- \sin \tau \right)^2
+\frac{1}{2}\left[\frac{\sin \tau}{k_0-\sin \tau}-
\frac{3}{2} \frac{\cos^2 \tau}{(k_0-\sin \tau)^2} \right].
\label{thetahill}
\end{equation}

Equation (\ref{hill}) is known as the Hill equation for parametrically driven system, and the function $\Theta(\tau)$ is called the excitation function. It is known \cite{Merkin} that the Hill equation can demonstrate the non-stationary behavior known as parametric resonance. The specific form of the instability regions in the parameters space can be found either from the numerical analysis of the Hill equation which implies its solution on a single period of the excitation function, or by the analytical approximation when the Hill equation is transformed to some other form with known areas of instability.
The possibility of such transformation crucially depends on the Fourier spectrum of the excitation function $\Theta(\tau)$. Our analysis of Fourier components of the function (\ref{thetahill}) has shown that it is the zeroth cosine harmonic and the first sine harmonic that dominate over the major part of our parameters which include weak and moderate fields. One may expect that the effects of higher harmonics appearing in cases when their amplitude grows up will tend to enlarge the instability regions found for the dominating lower harmonics.
As to the case when $k_0 \to 0$ in (\ref{thetahill}), it can be shown that one arrives to the excitation function where the single harmonic with double frequency $2\omega_0$ dominates, and the subsequent analysis is similar to the case presented below with the substitution $\omega_0 \to 2\omega_0$. 
Hence, for the prediction of the instability effects the Hill equation can be approximated by the equation where $\Theta(\tau)$ is replaced by the combination of its zeroth cosine and first sine harmonic. By making the phase shift $\tau \to \tau + \pi/2$ which changes $\sin \tau$ to $-\cos \tau$ we arrive to the following equation for $\sigma_1(\tau)$:

\begin{equation}
\frac{d^2 \sigma_1}{d \tau^2}+ \left(\delta+ \varepsilon \cos \tau \right) \sigma_1=0.
\label{Mathieu}
\end{equation} 

Equation (\ref{Mathieu}) is the Mathieu equation. This equation describes parametric resonance and has well-defined areas of stability and instability known as the Ince-Strutt diagram \cite{Merkin} in the plane of the parameters $(\delta, \varepsilon)$. The expressions for these parameters follow directly from the Fourier decomposition of the excitation function $\Theta(\tau)$:

\begin{equation}
\delta=\left(k_0^2+\frac{1}{2} \right)\frac{\Omega^2}{\omega_0^2}
+\frac{1}{4}\left(1-\frac{k_0}{\sqrt{k_0^2 - 1}} \right),
\label{delta}
\end{equation}

\begin{equation}
\varepsilon=2k_0 \frac{\Omega^2}{\omega_0^2}
+2k_0 \left(\frac{k_0}{\sqrt{k_0^2 - 1}}-1 \right)-\frac{3}{2\sqrt{k_0^2-1}}.
\label{epsilon}
\end{equation} 

The parameters $(\delta, \varepsilon)$ defined in (\ref{delta}), (\ref{epsilon})
via the values of $\mathcal{E}_0$ and $\overline{k_{y0}}$ determine the regimes of parametric oscillations. The parameter $\delta$ approaches constant non-zero value when the electric field amplitude $\mathcal{E}_0 \to 0$, and this value is defined by the initial value of $\overline{k_{y0}}$. 
For moderate fields the driving amplitude $\varepsilon$ behaves almost linearly with $\mathcal{E}_0$, as it follows from the definitions (\ref{k0}), (\ref{omelec}). According to the Ince-Strutt diagram, the onset of instability for growing $\varepsilon$ is possible as the parametric resonance, which corresponds to the development of irregular dynamics for mean value of spin variable $\overline{\sigma_x}$, and hence for the other coupled degrees of freedom in our system.

For our set of parameters (\ref{k0}), (\ref{omelec}) one can easily check that 
for the driving frequency $\omega_0 = 0.58 \cdot 10^{12}$ $s^{-1}$ considered below the amplitude of the driving field as low as $\mathcal{E}_0 \approx 1$ V/cm may put our system into an area of instability in the Ince-Strutt diagram.
When this or higher electric field is driving the dynamics, one may expect the development of irregular regimes of spin and coordinate dynamics which are coupled through the system (\ref{sysevol-a})- (\ref{sysevol-e}). In the next sections we will observe the justification of this finding by fully quantum mechanical approach and computational analysis for several values of driving field.

\section{Quantum dynamics and Floquet states}

In this and in the following Sections we perform the quantum mechanical treatment of the electron state evolution under the spatially uniform and time-periodic electric field described by the term (\ref{driving}) in the Hamiltonian (\ref{hfull}). We would like to see the resonance dynamics, so the frequency $\omega_0$ of the driving field matches the level splitting $(E_{n_0+1}-E_{n_0})/\hbar$ in the region of mostly populated levels by the initial wavepacket, which is the medium part of the spectrum $n_0 \approx 54$ (see Fig.\ref{figmodel}(d,e)).
As we have discussed in Section II, in the case of a multi-level mesoscopic QD with the length $L=3$ mkm the level splitting in the medium part of the spectrum is almost equidistant with the level spacing given by (\ref{den0}).
Such level splitting corresponds to the frequency
$\omega_0 = 0.58 \cdot 10^{12}$ $s^{-1}$ being in the sub-terahertz range (the so called W-band) for which one can consider the field as being quasistationary since $\hbar \omega_0$ is lower than the band gap by at least one order of magnitude. In this case the scalar potential $V(y,t)$ is described by a small overall amplitude on the scale of the mesoscopic QD, so one can introduce this potential into the envelope function Hamiltonian $H(y,t)=H_{\rm{QD}}+V(y,t)$.

The periodic driving allows us to apply some of the tools from the Floquet analysis \cite{Stockmann,Reichl,DIM,Rabi2012,billiard2013} for understanding
the system evolution. What is the most relevant for our system is the structure of the Floquet states where the $(s)$- th
eigenstate is written as a vector $A^{(s)}_n$ in the Hilbert space of the basis states $\Psi_n(y)$. These vectors are the eigenvectors of the one-period
propagator matrix $\mathbf{U} (T_0)$  where $T_0=2\pi / \omega_0$, which can be constructed from the evolution of the state $\Psi(y,t)$ in the basis
of states $\Psi_n(y)$,

\begin{equation}
\Psi(y,t)=\sum_n C_n(t) \Psi_n(y),
\label{psit}
\end{equation}

obeying the non-stationary Schr\"{o}dinger equation for the envelope function

\begin{equation}
i\hbar \frac{\partial \Psi}{\partial t}=(H_{\rm{QD}}+V(y,t)) \Psi,
\label{se}
\end{equation}

with the initial condition $C_n(0)=\delta_{n n_0}$ considered sequentially for all levels $n_0$ \cite{Reichl,DIM}.
The equation (\ref{se}) is transformed into a system of ordinary linear differential equations for coefficients $C_n(t)$ by projecting it
on the basis of the states $ \Psi_n(y)$, and this system is solved by standard numerical packages.
The eigenvalues of $\mathbf{U} (T_0)$ labeled by index $(s)$ have the form $\exp(-i E^{(s)}_Q T_0 / \hbar)$ where $E^{(s)}_Q$ are the corresponding
quasienergies.
It is known that the information contained in the quasienergy level spacing distribution can describe the regimes of the driven evolution
as regular or chaotic, depending on whether or not such distribution demonstrates the Poissonian or non-Poissonian behavior \cite{Stockmann,Reichl}.
In Fig.\ref{Floquet}(a) we show the level spacing distribution $\rho(\Delta E_Q)$ for three different driving strengths in (\ref{driving}),
$\mathcal{E}_0=0.2$ V/cm (dash-dotted curve), $\mathcal{E}_0=1$ V/cm (solid curve), and $\mathcal{E}_0=2$ V/cm (dotted curve).
Although the number of energy levels in our system is quite limited to see a full-developed smooth distribution,
it can be seen that for the weak driving the level statistics looks like the Poissonian one with the most of quasienergy levels grouped with
small spacing $\Delta E_Q$ of the order of $0.005$ meV. As the driving increases, the level statistics progressively transforms to a non-Poissonian type
with the maximum located near $0.03$ meV for  $\mathcal{E}_0=1$ V/cm and near $0.1$ meV for  $\mathcal{E}_0=2$ V/cm.
According to the basic concepts of quantum chaos \cite{Stockmann,Reichl}, this result can be viewed as an indication of the transition to chaos in
our system with increased amplitude of periodic driving.

It is useful to compare the properties of the quasienergy level statistics with the results for the quasiclassical dynamics obtained in the previous Section. For the weak driving amplitude $\mathcal{E}_0=0.2$ V/cm one has the following set of parameters (\ref{delta}), (\ref{epsilon}) for the Mathieu equation by considering the conservative estimate of $\overline{k_{y0}}=\pi/L$:
$\delta=3.92$, $\varepsilon=0.2$. According to the Ince-Strutt diagram \cite{Merkin}, this corresponds to the region of stability, which is reflected in the Poissonian type of statistics in Fig.\ref{Floquet}(a), being the sign of regular dynamics. Then, when the electric field amplitude is increased, we have
$\delta=4.08$, $\varepsilon=0.9$ for $\mathcal{E}_0=1$ V/cm, and
$\delta=4.4$, $\varepsilon=2.15$ for $\mathcal{E}_0=2$ V/cm, respectively.
Both these points in the $(\delta,\varepsilon)$ plane fall within the areas  of instability on the Ince-Strutt diagram, which means that the irregular dynamics is possible in the quasiclassical limit. These finding is justified by the full quantum mechanical treatment within the Floquet approach, being manifested in the transformation of the level statistics in Fig.\ref{Floquet}(a) from Poissonian to non-Poissonian type during the increase of the electric field amplitude.

Besides the quasienergy spectra, the structure of the Floquet eigenvectors $A^{(s)}_n$ can give a lot of information regarding the possibilities
of chaotic regimes for the evolution under periodic driving \cite{Stockmann,Reichl,DIM}. In particular, the presence of
the states which are extended in the Hilbert space formed by basis functions, that is,
described by high values of variance $\sigma_n$,

\begin{equation}
\sigma^{2}_n=\sum_{n}\left( n -{\bar n} \right)^2|A_{n}|^{2},
\label{sigman}
\end{equation}

where ${\bar n}=\sum_{n}n|A_{n}|^{2}$,
corresponds to the regimes of diffusion in the Hilbert space of the initial state along such extended
Floquet states, which can be viewed as the quantum counterpart of the classical chaos development. Hence, it is of interest to look
at the distribution for all of the quasienergy eigenstates in the $(\bar{n},\sigma_n)$ coordinates where $\bar{n}$ is the mean level number measuring
the center of
the Floquet state in the basis, and $\sigma_n$ is the variance, or width in the Hilbert space. In Fig.\ref{Floquet}(b) we plot the $(\bar{n},\sigma_n)$
distributions for the Floquet eigenstates for three different driving strengths, $\mathcal{E}_0=0.2$ V/cm (stars),
$\mathcal{E}_0=1$ V/cm (filled circles), and $\mathcal{E}_0=2$ V/cm (open circles). It is clear that the level variance $\sigma_n$
in general increases with the driving strength which is an expected effect (although a certain saturation with growing $\mathcal{E}_0$
is present), and the extended states with $\sigma_n \approx 32$ exist at moderate and strong driving, meaning the presence of the diffusion
in the Hilbert space into a substantial part of the spectrum totaling $108$ levels for the present set of model parameters.
We may see from Fig.\ref{Floquet}(b) that the difference between the quasienergy state statistics for $\mathcal{E}_0=1$ V/cm and $\mathcal{E}_0=2$ V/cm
is only quantitative since both fields correspond to the quasiclassically unstable regions,
so only moderate driving fields not exceeding the scale of
1 V/cm are required for the excitation of
the irregular or chaotic regimes in our system.
We thus can conclude that the analysis of the Floquet eigenstates demonstrates
the possibility of excitation of the diffusion regimes in the Hilbert space if the initial states are located in the region of maximum
variance $\sigma_n$ near the center of the spectrum.
In the next Section we will confirm this assumption by integration of the nonstationary Schr\"{o}dinger equation with Hamiltonian
$H_{\rm{qd}}+V(y,t)$ over the continuous time interval. The reason for such approach is that a substantial part of the evolution
takes place between the stroboscopic moments of time $T_n=n T_0$ which are in the focus of the Floquet stroboscopic approach.
To obtain more detailed picture, we proceed with direct numerical integration for continuous time with suitable time grid catching all of
the essential details of the dynamics, and providing also a perfect match between the continuous and Floquet approaches.

\section{Evolution in the clean limit}

We begin with the analysis of the driven evolution of the wide wavepacket (see Fig.\ref{figmodel}(c),(d))
which is described by the Schr\"{o}dinger equation (\ref{se}) with moderate driving amplitude $\mathcal{E}_0=1$ V/cm
of the driving electric field (\ref{driving}).
The initial state $C_n(0)$ occupies a narrow part of the Hilbert space near the center of the spectrum,
as it can be seen in  Fig.\ref{figmodel}(d). We solve the equations of motion for $C_n(t)$ from several hundreds
to several thousands of periods $T_0=2\pi/\omega_0$ which is the unit of time in our model, where $\hbar \omega_0$ is the spacing
between a selected pair of levels near the center of the spectrum.

The initial state of the wavepacket injected from the left barrier
into the QD is characterized by the spin polarization in units of $\hbar/2$ as $S_x=1$, $S_y=S_z=0$. As we have mentioned in Section II, for all the basis states (\ref{psileft})-(\ref{psidot}) the mean spin is always in the plane of the 2D TI,  that is, $S_z=0$. However, if a time-dependent mixture (\ref{psit}) of such states is considered to be formed by the initial packet
 or by the non-stationary driving $V(y,t)$, the resulting spinor wavefunction may correspond to the state where the out-of-plane $S_z$ spin component is present. We will discuss it in detail below.

We look at the evolution of the quantum mechanical mean values, or observables, for several variables of interest, both for coordinate and spin degrees of freedom. As it is known from the classical mechanics, the evolution of the driven system can be represented in terms of canonically conjugated variables such as $(x_i,dx_i/dt)$ shown in phase plots. For a quantum mechanical system the concept of trajectories is not directly available, and
one can consider the dynamics of mean values of such variables.
The velocity operator introduced as ${d x_i/dt}=\frac{i}{\hbar} \left[H, x_i \right]$ gives for $v_y=dy/dt$ and the Hamiltonian (\ref{hqd}) the following form:

\begin{equation}
v_y=v_F \sigma_z
\label{vy}
\end{equation}

Note that the time dependence of its mean value is defined by Eq.(\ref{sysevol-a}) obtained for  the quasiclassical dynamics.
This result means that the velocity is effectively represented on the "phase space plot" by the $z$ component of spin,
so the first pair of the mean values to be plotted is $(y,S_z)$. Hereafter for brevity we omit the $\overline{(\ldots)}$ mark for mean values to be plotted on figures.
For our model of coordinate and spin dynamics it means that these two channels are tightly coupled from the very beginning,
and we may expect certain common characteristics of their evolution, as it was already shown for semiconductor mesoscopic QD
with spin-orbit coupling \cite{billiard2013}.

The general expressions which allow us to calculate the mean values of coordinate and velocity for the state (\ref{psit}) via the matrix elements $y_{ij}$ of the position operator $y$ are

\begin{equation}
\overline{y}(t) =\sum_{ij} C_i^*(t) C_j(t) y_{ij},
\label{ymean}
\end{equation}

\begin{equation}
\overline{v_y}(t) =\frac{i}{\hbar} \sum_{ij} C_i^*(t) C_j(t) \left(E_i - E_j \right)y_{ij}.
\label{vymean}
\end{equation}

One can see that the velocity mean value (\ref{vymean}) is directly related not only to the spin via (\ref{vy}) but equivalently
to the position operator $y$ since its matrix elements define both expressions (\ref{ymean}) and (\ref{vymean}) together with the energy levels structure.
By comparing two approaches (\ref{vy}) and (\ref{vymean}) for the velocity operator one can see that in the present model the spin projection indeed plays the role of the momentum for the classical spinless oscillator with finite mass. Hence, one may expect that the phase plots for the coupled coordinate and spin dynamics may resemble to some extent the conventional phase plots for the driven oscillator plotted for the $(y, v_y)$ variables.
It can be mentioned that for strictly equidistant spectrum $E_i - E_j=\hbar \omega_0$ from (\ref{ymean}) and (\ref{vymean}) one gets $\overline{v_y}=i \omega_0 \overline{y}$ just as for the linear oscillator.
In our model, however, the level spacing is not purely constant, and such simple relation is only approximate but not exact.

The second pair of variables to be plotted together is the in-plane spin projection represented by the mean values of $(S_x,S_y)$ spin components.
This choice is motivated by the inherent structure of the Hamiltonian (\ref{hqd}).
Namely, the internal part of the QD region is described by the Weyl
Hamiltonian coupling the $(y,S_z)$ degrees of freedom, and the surrounding barriers are polarized in the $(S_x,S_y)$ plane which couples also these two spin components to the other degrees of freedom. As a result, the spin vector is subject to evolution for all of its projections which are coupled to the one-dimensional spatial motion along the y-direction in the QD.
Such pairs of spin variables have been considered in several studies on the spin-resolved systems \cite{KS2009,billiard2013,Yang2007},
and is convenient in representing, for example, the in-plane spin precession.

For the numerical calculations we consider the time interval of $400$ periods of driving field with $200$ points per period for the graphical representation.
These parameters cover both the time span needful for the stationary regime of the dynamics to be established,
and the time grid which catches the significant non-vanishing Fourier components of the evolution of observables.

In Fig.\ref{figdrivencleanwide} we show the results for the driven evolution for the initial state represented by the wide
wavepacket from Fig.\ref{figmodel}(c) with zero mean value of the quasimomentum,
$\overline{k_{y}}(0)=0$.
The initial point at $t=0$ is marked as the black circle "A".
Panel (a) shows the "phase space" plot of evolution in the $(y,S_z)$ coordinates and panel (b) shows the in-plane spin precession in the $(S_x,S_y)$ plane. In Fig.\ref{figdrivencleanwide}(c) the evolution of the variance of the level number $\sigma_n(t)$ is shown which describes the spreading of the initial state $C_n(0)$ in the Hilbert space of the basis states.
Also, we are interested in the dynamics of the variance $\sigma_y(t)$ for packet width in the coordinate space,

\begin{equation}
\sigma^{2}_y(t)=\overline{\left(y- \overline{y}(t) \right)^2}.
\label{sigmay}
\end{equation}

The time dependence for this quantity is shown in Fig.\ref{figdrivencleanwide}(d). It is also of interest to look at the spatial distributions along the QD for the charge density $\rho(y,t_0)$ and some of the spin density components $S_i(y,t_0)$, $i=x,y,z$, at certain moments of time. The spatial profiles of charge and spin density help in understanding on which spatial scale one can measure the charge and spin spots in actual experimental setups. We present an example of charge and spin densities plotted at the specific moment of time in Fig.\ref{figdrivencleanwide}(e).

Besides tracking the evolution of the mean values, we are interested in their Fourier power spectra

\begin{equation}
I_{\xi}(\omega)=\left| \int_{-\infty}^{+\infty} \xi(t) e^{-i \omega t} dt \right|^2,
\label{powerspec}
\end{equation}

where $\xi$ is the variable of interest.
Since we have obviously considered large, but finite intervals of time,
the Fourier power spectra (\ref{powerspec}) were calculated by the Fast Fourier Transform with limits of time actually used in our simulations of dynamics.
In panels (f),(g) and (h) in Fig.\ref{figdrivencleanwide}  we show the Fourier power spectra for the variables $y$, $S_z$ and $S_x$, respectively.

Let us discuss now the meaning of the results presented in Fig.\ref{figdrivencleanwide}.
First, we take a look onto the trajectories in the space of $(y,S_z)$ and $(S_x,S_y)$ pair of variables shown in Fig.\ref{figdrivencleanwide}(a),(b).
One can see that the regular trajectories for these phase plots are accompanied
by the surrounding areas of the "chaotic sea", although the general oscillating character of the wavepacket evolution is still visible.
We can say that the phase portrait in Fig.\ref{figdrivencleanwide}(a) in general resembles the phase trajectories of a driven classical
oscillator in the irregular regime of dynamics.
The onset of irregular motion is further pronounced in the in-plane spin dynamics in Fig.\ref{figdrivencleanwide}(b).
In general terms, one can state that the spin evolution becomes largely irregular.
As it was found in Section III, the quasiclassical spin dynamics for the driving field amplitude $\mathcal{E}_0=1$ V/cm corresponds to the unstable region of the Mathieu equation (\ref{Mathieu}). We see that the full quantum mechanical treatment leads to the same conclusion about the onset of chaotic dynamics for the electric field amplitude and frequency corresponding to quasiclassically unstable area. 
Another justification for the correspondence between the full quantum mechanical and quasiclassical treatment is the initial wavepacket width considered here. In this Section it corresponds to a narrow distribution in the Hilbert space of the basis states, which makes the quasiclassical approach applicable, and the main conclusions from the quasiclassical and full quantum approaches support each other.

As to the spin dynamics in general, one may describe it as a combination of precessions with generally incommensurable frequencies around the directions of effective magnetic fields stemming from the Hamiltonian (\ref{hqd}),
namely, the z-oriented $k_y$-dependent field inside the QD and the x-oriented barrier field. As a result, the spin dynamics becomes rather complex. A large clustering area near the origin for the in-plane components  $(S_x,S_y)$
reflects the faster frequency of precession around the z-aligned effective magnetic field inside the QD where the wavefunction is mostly located, resulting in the averaged in-plane spin components $(S_x,S_y)$ being close to zero, as we can see in Fig.\ref{figdrivencleanwide}(b). The observed spin precession can be considered as typical for systems with strong spin-orbit coupling, which was found, for example, for the models of spin dynamics in semiconductor quantum dots \cite{KS2009,Rabi2012}.

The concept of irregular dynamics or chaos development can be supported by analysis of the driven evolution in the Hilbert space
of basis states. Here the onset of chaos usually corresponds to the growth in time of the number of energy levels involved into evolution,
which is sometimes called as quantum Arnol'd diffusion \cite{DIM}. Our Floquet analysis of the quasienergy eigenfunctions in Section IV indicates that the periodic driving with amplitudes of $1\ldots 2$ V/cm may induce the formation of the Floquet states which are deeply extended into a substantial part of the energy spectrum, see Fig.\ref{Floquet}. Thus, we may expect the variance $\sigma_n$ measuring the number of levels involved into evolution to be as high as the maximum number reached by the Floquet states. This assumption is confirmed by the plot of $\sigma_n(t)$ in Fig.\ref{figdrivencleanwide}(c) where almost linear growth of level number is present at the initial stage of evolution where the quantum-classical correspondence is mostly pronounced \cite{Gutzwiller,Stockmann,Reichl}. Such growth is usually attributed to the onset of chaotic dynamics, or diffusion in Hilbert space, which provides another correspondence between the findings on the instability regions within the quasiclassical approach in Section III and full quantum mechanical treatment. After some time, however, the discrete character of the quantum mechanical spectrum of a finite motion inside QD leads to the saturation of the level number involved in the evolution, and the diffusion in the Hilbert space effectively stops \cite{DIM}. This can be seen in  Fig.\ref{figdrivencleanwide}(c) where the linear growth of $\sigma_n$ transforms into oscillations with stable mean value. We can say that the chaotic behavior in our quantum system has a transient nature.

As to the dynamics in the real space inside the QD, the evolution of the packet half-width is presented in Fig.\ref{figdrivencleanwide}(d).
The packet width essentially does not grow with time, and the packet at each moment of time occupies effectively only a limited area inside the QD. This finding is illustrated by an example of the spin and charge density distributions inside the QD shown for $t=395 T_0$ in Fig.\ref{figdrivencleanwide}(e) as solid and dashed lines, respectively.
One may see that the packet occupies a substantial part of the QD, however,
its effective width has the value close to the width of the initial packet, see Fig.\ref{figmodel}(c), wavepacket (1).
As we have mentioned earlier, such stable behavior of the packet width during the driven evolution can be attributed to the nearly equidistant character of the energy levels of the system which can trigger certain properties of the coherent states in the driven evolution.

The manifestation of chaotic or at least strongly irregular regimes for the driven evolution is supported by the Fourier power spectra
for the coordinate and spin observables plotted in Fig.\ref{figdrivencleanwide}(f)-(h). One may see that the driving induces a large number of harmonics of driving frequency $\omega_0$ both for coordinate and spin, especially the in-plane component $S_x$ (and similarly for $S_y$ which
is not shown here).
The presence of large number of harmonics is a strong indication of irregular dynamics \cite{Gutzwiller,KS2009,billiard2013} which supports the quasiclassical results from Section III on the onset of unstable dynamics for the considered amplitude of the driving field.

Another possible manifestation of chaos is the presence of positive Lyapunov exponents \cite{Gutzwiller,Reichl}
which measure the rate of divergence of two initially close trajectories in the phase space,

\begin{equation}
\lambda = \rm{lim}_{t\to \infty} \frac{1}{t} \log \frac{d(t)}{d(0)},
\label{Lyapunov}
\end{equation}

where $d(t)$ and $d(0)$ are the present and initial distances, respectively.
The infinite limit in (\ref{Lyapunov}) can be tracked also by a continuously monitoring with growing time where $\lambda=\lambda(t)$
tracks local transition between regular and chaotic regimes.
In Fig.\ref{figdrivencleanwide}(i) we plot the dependence of $\lambda(t)$ for two initially close wavepackets which mean values of coordinate are
shifted slightly along $y$ at $t=0$.
One can see that at the beginning of the evolution the region with positive $\lambda(t)$ indeed exists which corresponds to the linear growth
of the level number $\sigma_n$ involved into the dynamics, see Fig.\ref{figdrivencleanwide}(c). Both these plots support the presence of chaotic
dynamics at the initial stage of the evolution when the discrete character of the quantum spectrum has not yet manifested itself so much.
After the initial transient period the evolution tends to transform to a quasi-regular regime with the stable number of levels involved into dynamics,
and the Lyapunov exponent reduces to zero, as in can be seen in Fig.\ref{figdrivencleanwide}(i). It should be noted that such behavior
is known in quantum systems with irregular dynamics \cite{Gutzwiller,Stockmann,Reichl,DIM}. However, the results obtained there
were mainly for the spinless systems with quadratic spectrum having a certain classical analogue.

To conclude this Section, we can state that even in a quantum system which lacks the classical analogue such as the system with
Hamiltonian (\ref{hqd}) one can observe certain traits of the development of irregular phenomena which are present in classically chaotic systems and are in good agreement with quasiclassical treatment. It should be stressed that
such effects may arise in considered structures at driving fields as low as several V/cm. This means that the apart the fundamental questions on the degree of irregularity of the electron and spin dynamics
in the systems with strong spin-orbit coupling our findings can also be important for the nanodevice designers and experimentators
for future applications of the TI-based structures.

\section{Evolution in the presence of a disorder}

The presence of some sort of disorder in the form of spatially non-uniform potential at the TI edge or the potential caused by the defects is inevitable in any real structure and should be addressed in the problem of the electron evolution. In this Section we insert an additional stationary disorder potential of the form

\begin{equation}
U_d(y)=U_0 f(y)
\label{disorder}
\end{equation}

into the right-hand side of the non-stationary Schr\"{o}dinger equation (\ref{se}) where the potential amplitude $U_0$ is multiplied by a
random function $f(y)$ described by a uniform random distribution from $0$ to $1$ along the QD where $0 \le y \le L$.
It is known that the presence of disorder alone with the potential that preserves the time reversal invariance such as the scalar potential (\ref{disorder}) does not break the topological protection of the edge states.
For the QD considered in our model time reversal symmetry has already
been broken by the presence of magnetic barriers, so the disorder potential may
induce, for example, the transitions between the states with different spin
polarization.

The interest to the influence of external disorder potential on the wavepacket evolution in materials such as Dirac fermion materials started to arise during the last years, leading to sometimes unexpected results. For example, it was shown recently that the inclusion of a static 1D disorder potential into the model of wavepacket propagation in graphene and related Dirac fermion materials may cause the so called electron supercollimation, i.e. effect when
the wavepacket moves undistorted along certain direction \cite{Choi2014}. So, it is a challenging and intriguing task to consider the effects of the disorder potential on the driven dynamics in our model of QD in TI.

The matrix elements of $U_d(y)$ from (\ref{disorder}) contribute to the dynamics of the coefficients $C_n(t)$ for the wavefunction (\ref{psit})
together with the ones from the driving term $V(y,t)$. We consider an example of  the amplitude $U_0=0.5$ meV which is comparable with typical
interlevel distance (\ref{den0}) equal to $0.38$ meV, i.e. we insert a moderate disorder. This can be justified by a typical high quality and high mobility of samples usually fabricated and studied in the experiments with TI \cite{TI},
which have the mean free path of the order of the QD length L, and low temperatures of around or below 1 K which produces
the level broadening of the order of $0.05$ meV.
For the initial condition we put a narrow wavepacket,
see the packet profile shown by curve (2) in Fig.\ref{figmodel}(c).
We take the same driving amplitude $\mathcal{E}_0=1$ V/cm,
and the same other parameters as in the previous Sec.

One can expect certain modifications of the evolution for both coordinate and spin degrees of freedom, when the disorder amplitude $U_0$ exceeds the energy of the driving field. In Fig.\ref{figr05narrow} we show the evolution of the narrow wavepacket under the driving with $\mathcal{E}_0=1$ V/cm,
with a disorder amplitude $U_0=0.5$ meV which exceeds the typical energy of the driving field $e\mathcal{E}_0 L=0.3$ meV.
The numbering of the figure panels is similar to Fig.\ref{figdrivencleanwide} from the previous Section. One can see that the inclusion of disorder leads to more uniformly distributed trajectories in the phase space
of the $(y,S_z)$ variables shown in Fig.\ref{figr05narrow}(a).
We can say that the disorder reduces the degree of correspondence with phase plots for the classical driven oscillator.
The in-plane spin components $(S_x,S_y)$ demonstrate a more pronounced tendency
to cluster near the coordinate origin $(0,0)$ with growing time meaning that the spin precession here is accompanied by collisions of the electron with the inhomogeneities of the potential (\ref{disorder}) and leading to the effective spin relaxation.
As to the off-plane spin component $S_z$, it still demonstrates a full-scale oscillating behavior representing the electron velocity in our model,
but within a well-established chaotic sea visible in the $(y,S_z)$ plot.
The disorder leads to an interesting effect on the number of levels $\sigma_n$ effectively involved in the evolution which is shown in Fig.\ref{figr05narrow}(c). Starting from the initially high number of basis states present in the decomposition of a narrow wavepacket, this number begins to decrease progressively, with the average level number (not shown)
moving down from the Dirac point. Such a form of localization in the Hilbert space can be viewed as the decrease of the irregularity when the dynamics of the system becomes in fact more regular in terms of the number of states involved into evolution. We can say that the presence of static disorder inhibits the development of dynamical chaos, although it does not suppress it completely.

The variance of the packet half-width shown in Fig.\ref{figr05narrow}(d) shows the oscillating behavior with saturating amplitude
which again demonstrates the effect of the wavepacket maximum width stability induced by the periodic driving, which is maintained
even in the presence of a strong disorder. An example of the charge  $\rho(y)$ and spin density $S_z(y)$ distributions
at the end of the observation frame $t=395 T_0$ shown in Fig.\ref{figr05narrow}(e) supports this finding, demonstrating
 a well-localized packet near one of the edges of the QD.
These findings regarding the dynamics are supported by the Fourier power spectra
shown in Fig.\ref{figr05narrow}(f)-(h) where the spin components demonstrate a more quickly vanishing spectra compared to the coordinate
one. This can be attributed to the effects of the spin precession in the presence of the collisions caused by disorder.
One can say that the observed effects of the disorder to some extent lead to the reduction of the dynamical chaos. This observation is important for further experimental and technological applications since in real structures the presence of some degree of disorder is inevitable, and in certain cases it can play a positive role as a damper of chaotic regimes of the dynamics.

To conclude this Section, we can say that by comparing Fig.\ref{figdrivencleanwide} and Fig.\ref{figr05narrow} one may see that the principal features of almost all respective panels on these figures look similar. In contrast, the initial wavepacket in Fig.\ref{figdrivencleanwide} is described by the quasiclassically-treatable narrow distribution in the Hilbert space of basis states, while the initial wavepacket in Fig.\ref{figr05narrow} is described by the wide distribution for which only the full quantum simulation is required. This similarity may be viewed as additional argument supporting the applicability of the quasiclassical approach derived in Section III since the full quantum treatment leads to the principally close results for both regions with or without the possibility to apply the quasiclassical method.

\section{Conclusions}

We have studied the dynamics of Dirac-Weyl wavepackets driven by periodic electric field in a mesoscopic quantum dot formed at the edge
of two-dimensional HgTe/CdTe topological insulator, where the motion of carriers is less sensitive to disorder and impurity potentials.
It was found that the presence of strongly coupled spin and charge degrees
of freedom in such driven system leads to the regimes of transiently irregular dynamics both in the clean limit and in the presence of the disorder. The quasiclassical analysis of spin dynamics allowed to find analytically the border between the regular and irregular regimes defined by the amplitude and frequency of the driving field in the framework of Mathieu equation, and was supported by the full quantum mechanical treatment.
The predicted onset of irregular regimes both in coordinate and spin channels which occurs for a mesoscopic quantum dot at the amplitudes of driving fields being as low as $1 \ldots 2$ V/cm is, in our opinion, an important feature of the considered structures both from fundamental and device point of view. The observed effects of disorder can be described in general as damping of the chaotic regimes of dynamics which is also important for possible experiments in real structures.
We believe that our findings are not limited to 1D edges of 2D topological insulators based in HgTe/CdTe quantum wells but also for the other systems with Dirac-Weyl spectrum which allows considering them as being rather general.
Apart from the basic questions on the degree of irregularity of the electron and spin dynamics in the systems with strong spin-orbit coupling our findings can also be taken into consideration by the nanodevice designers and experimentators
who plan to use the topological insulator-based structures for transport and information processing purposes, where the manifestations of irregularity
even at low driving fields may seriously affect their operational capabilities.

\section*{Acknowledgments}

This paper is dedicated to the memory of Prof. V.Ya. Demikhovskii who inspired much of our research, and who passed away on 01 May 2016. 
The authors are grateful to A.I. Malyshev, A.M. Satanin, E.Ya. Sherman, M.V. Ivanchenko, and S. Denisov for stimulating discussions.
The work is supported by the RFBR Grants No. 15-02-04028-a, 15-42-02254-r-povolzhye-a, 16-07-01102-a, 16-32-00683-mol-a, 16-32-00712-mol-a, and 16-57-51045-NIF-a, and is partially supported by the State Project of the Ministry of Science and Education RF No. 3.285.2014/K.

\newpage
\section*{Figure captions}

{\bf Fig.1}.
(a) Schematic view of a 1D quantum dot (gray strip labeled as QD) with length $L$ formed by two magnetic barriers $M_0$and $M_L$ at the edge of 2D topological insulator in HgTe/CdTe quantum well. The dot length $L=3$ mkm, the barrier height $M_0=M_L=E_g/2$ where $E_g=40$ meV is the band gap for the host material observed for typical HgTe/CdTe quantum well samples, and the arrows inside the barriers represent their polarizations actually considered in our model.
(b) Schematic representation of the discrete energy levels inside the QD with
the interlevel distance shown as a guide to the eye and not to scale.
The linear dispersion branches are shown for the free Weyl Hamiltonian (\ref{hweyl}) describing the edge states \cite{TI}, with corresponding spin mean values $S_z$. The boundaries of the bulk energy gap $E_g$ are plotted above and below as horizontal lines.
(c) Probability density distribution for Gaussian spin-polarized wavepackets with different widths and center locations representing the initial condition for the dynamics of the electron injected into the QD shown for the wide
(solid line (1)) and narrow (dashed line (2)) wavepackets.
(d),(e) Distribution of the expansion coefficients $|C_n|^2$ of the initial states from fig. (c) in the space of basis states for (d) the wide initial packet (1) and (e) the narrow initial packet (2), respectively.

\newpage
{\bf Fig.2}.
(a)  Level spacing distribution $\rho(\Delta E_Q)$ for the quasienergy levels for three different driving strengths $\mathcal{E}_0=0.2$ V/cm (dash-dotted curve), $\mathcal{E}_0=1$ V/cm (solid curve), and $\mathcal{E}_0=2$ V/cm (dotted curve).
For the weak driving $\mathcal{E}_0=0.2$ V/cm the level statistics is of the Poissonian type which corresponds to the parameter area of the stable dynamics for the Mathieu equation (\ref{Mathieu}) obtained within the quasiclassical approach. For stronger driving $\mathcal{E}_0=1$ V/cm and $\mathcal{E}_0=2$ V/cm the level statistics transforms to a non-Poissonian type indicating the irregular regime of quantum dynamics which corresponds to the unstable region of the quasiclassical approach.
(b) Distribution of the Floquet quasienergy eigenstates in the $(\bar{n},\sigma_n)$ coordinates where $\bar{n}$
is the mean level number measuring the center of the Floquet state in the basis, and $\sigma_n$ is the variance (width) in the Hilbert space,
for driving strengths $\mathcal{E}_0=0.2$ V/cm (stars), $\mathcal{E}_0=1$ V/cm (filled circles), and $\mathcal{E}_0=2$ V/cm (open circles).
The level variance $\sigma_n$ in general increases with the driving strength, and the extended states with $\sigma_n \approx 32$ exist at moderate and strong driving, meaning the presence of the diffusion in the Hilbert space into a substantial part of the spectrum and reflecting the possibility of irregular, or chaotic dynamics.

\begin{figure}[tbp]
\centering
\includegraphics*[width=170mm]{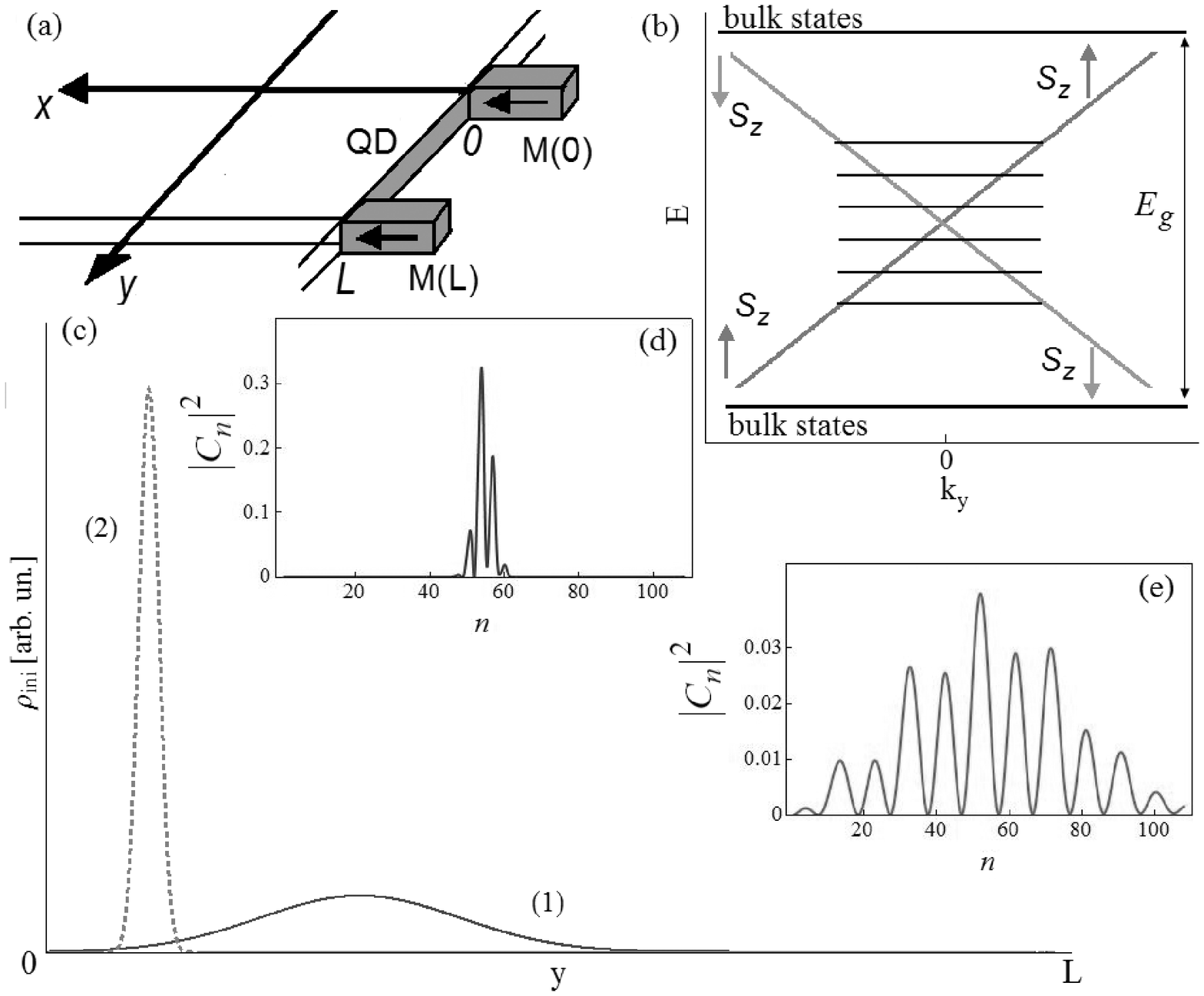}
\caption{}
\label{figmodel}
\end{figure}

\begin{figure}[tbp]
\centering
\includegraphics*[width=170mm]{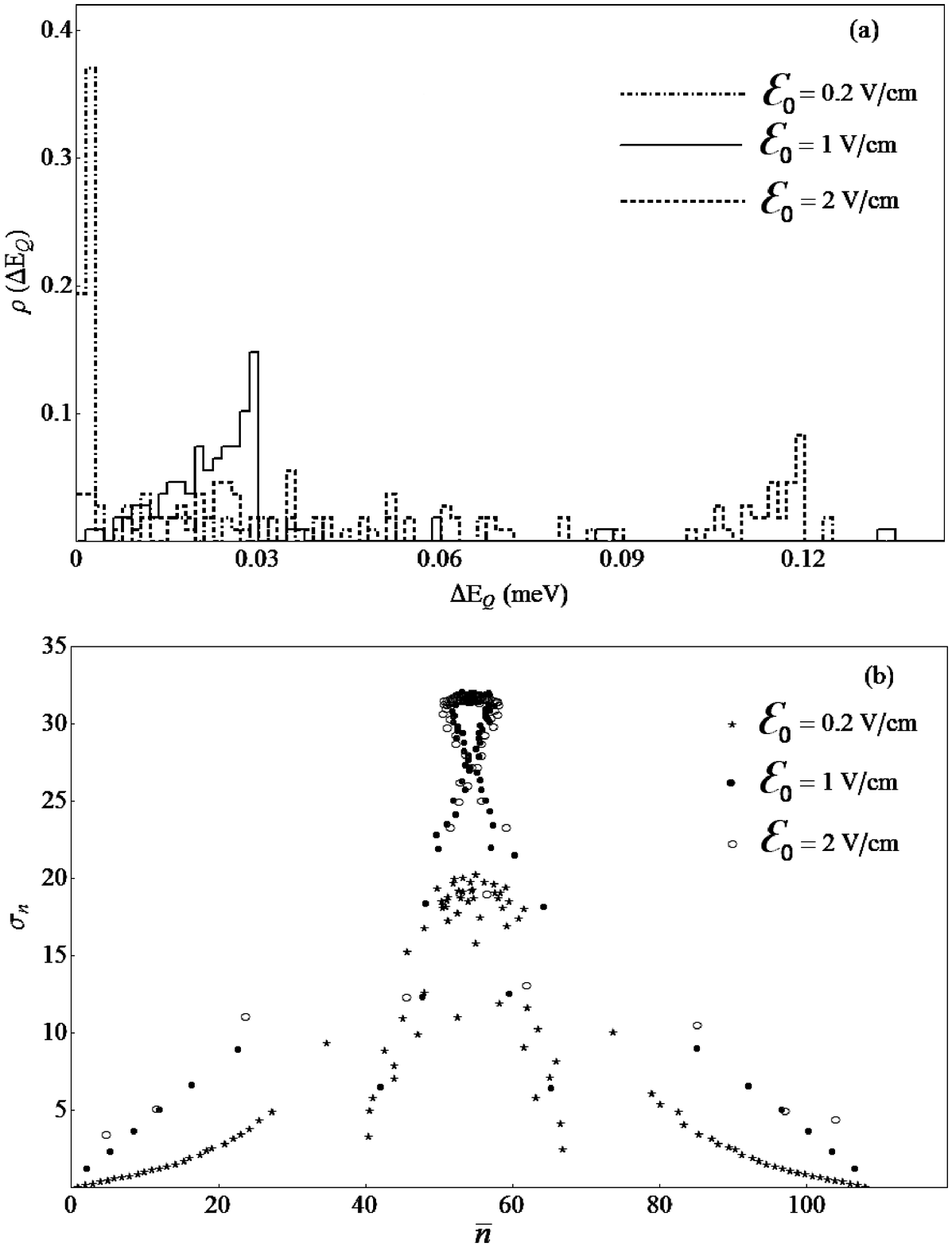}
\caption{}
\label{Floquet}
\end{figure}

\newpage
{\bf Fig.3}.
(a) Evolution in the "phase space" of the mean values $(y,S_z)$ representing classical coordinate and velocity, for the amplitude $\mathcal{E}_0=1$ V/cm of the driving electric field;
(b) Evolution of the in-plane spin components $(S_x,S_y)$. The initial point at $t=0$ is shown as black circle marked by "A".
The in-plane spin components demonstrate the tendency of clustering near the zero values with growing time which reflects the faster frequency of precession around the z-aligned effective magnetic field inside the QD where the
wavefunction is mostly located, resulting in the averaged in-plane spin components $(S_x,S_y)$ being close to zero.
(c) The evolution in the Hilbert space plotted as a number of levels effectively participating in the dynamics shows the linear growth at the beginning of the evolution which corresponds to the chaotic regime.
Later the level number saturates near a stable value corresponding to the width of the Floquet states in Fig.\ref{Floquet}.
(d) The packet half-width variance describing the spreading of the wavepacket in the real space inside the QD.
The initial half-width (see Fig.\ref{figmodel}(c), packet (1)) does not grow with time, and, as for the free evolution, at certain moments of time
the packet is narrowed.
(e) Charge density $\rho(y)$ (dashed curve)  and the $S_z(y)$ component of spin density (solid curve)
inside the QD taken at specific moment of time $t=395 T_0$. The packet is spraded along the QD, but the major part of the charge and spin density are concentrated on the width comparable with the width of the initial packet.
(f),(g),(h) Fourier power spectra (\ref{powerspec}) for the coordinate $y$ and two spin components $S_x$ and $S_z$.
Large number of harmonics of the driving frequency indicates the onset of strongly irregular motion which supports the quasiclassical results from Section III on the onset of unstable dynamics for the considered amplitude of the driving field.
(i) Evolution of the Lyapunov exponent (\ref{Lyapunov}) for two initially close wavepackets. At the beginning of the evolution
this exponent takes also positive values indicating the presence of chaotic regime, and later it decreases to zero when
the quasi-regular quantum dynamics is established.

\begin{figure}[tbp]
\centering
\includegraphics*[width=170mm]{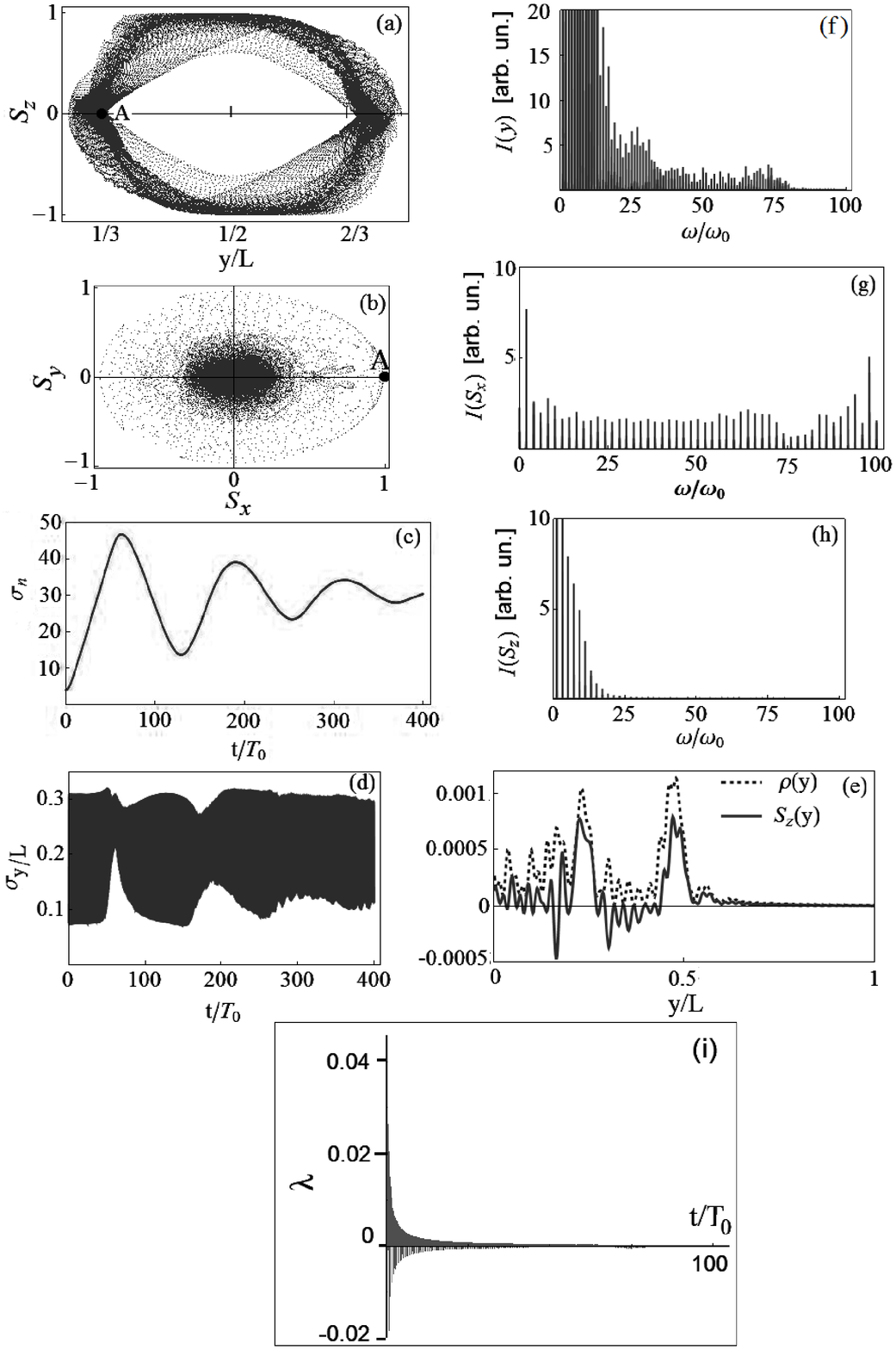}
\caption{}
\label{figdrivencleanwide}
\end{figure}

\begin{figure}[tbp]
\centering
\includegraphics*[width=170mm]{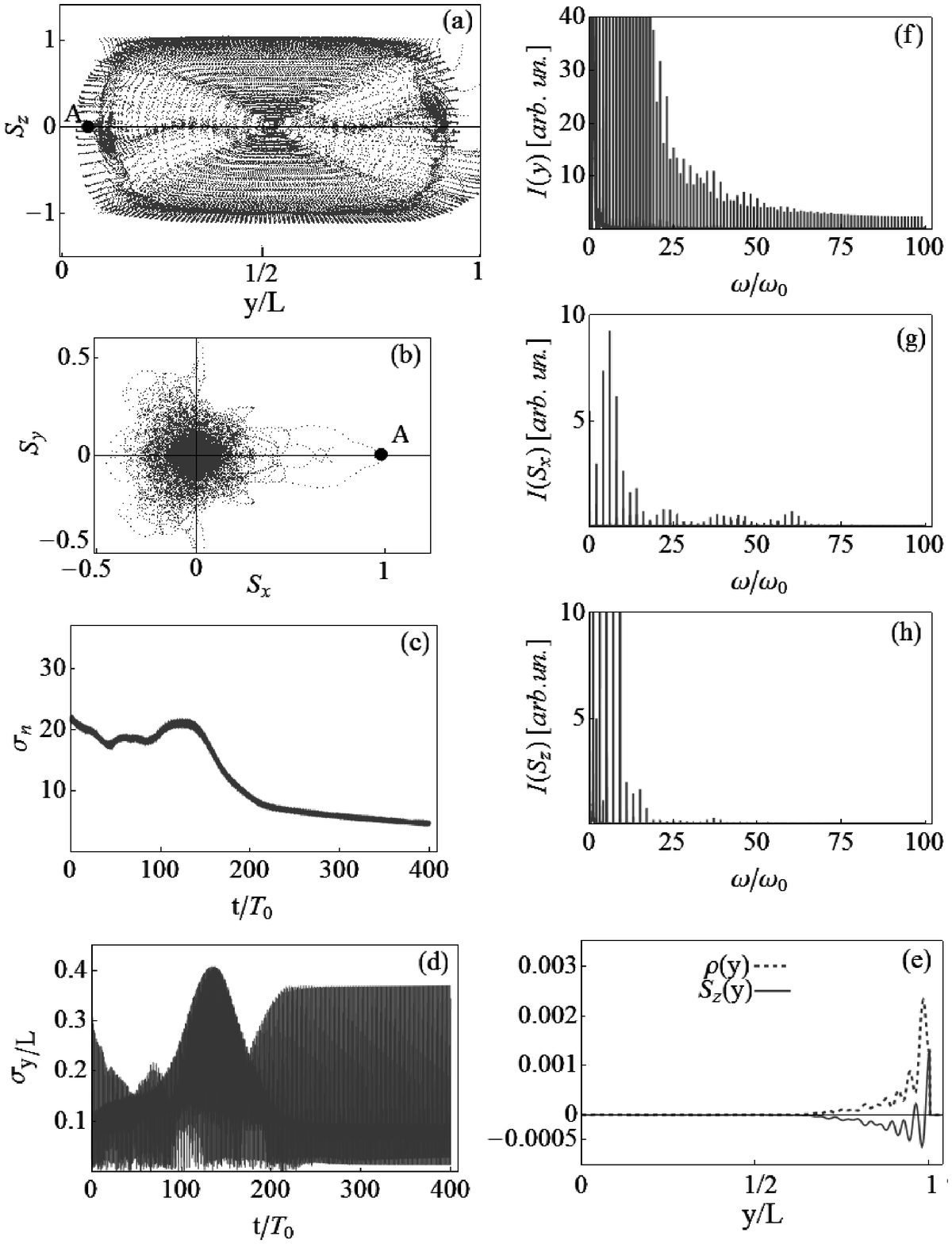}
\caption{}
\label{figr05narrow}
\end{figure}

\newpage
{\bf Fig.4}.
Same as in Fig.\ref{figdrivencleanwide} but for a narrow wavepacket (see Fig.\ref{figmodel}(c),(e))
taken as initial condition, and in the presence of the disorder potential (\ref{disorder}) with amplitude $U_0=0.5$ meV.
(a),(b) Phase plots for coordinate and spin mean values shown for pairs $(y,S_z)$ and $(S_x,S_y)$, respectively. The in-plane spin components show an enhanced tendency of clustering near the zero values with growing time.
(c) Evolution in the Hilbert space of $\sigma_n$ demonstrates the decreasing number of basis states participating in the evolution which can be viewed as an example of localization in the Hilbert space.
(d) The packet half-width dynamics shows the amplitude stability on long times
(e) An example of charge and spin density for $t=395 T_0$ demonstrates a well-localized packet even at the presence of strong disorder potential.
(f),(g),(h) Fourier power spectra for the coordinate $y$ and two spin components $S_x$ and $S_z$ showing the behavior where the disorder can enhance the in-plane spin relaxation.

\newpage

\newpage

\newpage

\newpage


\newpage
\begin{thebibliography}{99}

\bibitem{VolovikBook}
G.E. Volovik, The Universe in a Helium Droplet, Clarendon Press, Oxford, 2003.

\bibitem{Pal}
P.B. Pal, Am. J. Phys. 79, 485 (2011).

\bibitem{Lan}
Z. Lan, N. Goldman, A. Bermudez, et al, Phys. Rev. B {\bf 84}, 165115 (2011).

\bibitem{Vafek}
O. Vafek and A. Vishwanath, Annual Review of Condensed Matter Physics {\bf 5}, 83 (2014).

\bibitem{Wan}
X. Wan, A.M. Turner, A. Vishwanath, et al, Phys. Rev. B {\bf 83}, 205101 (2011).

\bibitem{Beenakker}
C.W. Beenakker, Rev. Mod. Phys. {\bf 80}, 1337 (2008).


\bibitem{TI} B.A. Bernevig, Topological Insulators and Topological Superconductors, Princeton University Press, Princeton, 2013;
S.Q. Shen, Topological insulators. Dirac equation in Condensed Matter, Springer Series in Solid-State Science, Springer-Verlag
Berlin Heidelberg, 2012; M.Z. Hasan and C.L. Kane, Rev. Mod. Phys 82, 3045 (2010); X-L Qi and  S-C Zhang, Rev. Mod. Phys. 83, 1057 (2011).

\bibitem{Volkov2009}
V.A. Volkov and I.V. Zagorodnev, Low Temperature Physics {\bf 35}, 2 (2009).

\bibitem{Gutzwiller}
M.C. Gutzwiller, Chaos in Classical and Quantum Mechanics, Springer-Verlag, New York, 1990.

\bibitem{Reichl}
L.E. Reichl, The Transition to Chaos. Conservative Classical Systems and Quantum Manifestations, 2nd Ed., Springer-Verlag, New York, 2004.

\bibitem{Stockmann}
H-J. St{\" o}ckmann, Quantum Chaos: An Introduction, Cambridge University Press, 1999.

\bibitem{Nakamura}
K. Nakamura and T. Harayama, Quantum Chaos and Quantum Dots, Oxford University Press, New York, 2004.

\bibitem{DIM}
V.Ya. Demikhovskii, F.M. Izrailev, and A.I. Malyshev,
Phys. Rev. Lett. 88, 154101 (2002); Phys. Rev. E 66, 036211 (2002);
A.I. Malyshev and L.A. Chizhova, J. Exp. Theor. Phys. 110, 837 (2010).

\bibitem{Zutic}
I. Z\v{u}ti\'{c}, J. Fabian, and S. Das Sarma, Rev. Mod. Phys. 76, 323 (2004).

\bibitem{Dyakonov}
Spin physics in semiconductors, ed. by M.I. Dyakonov, Springer-Verlag Berlin Heidelberg, 2008.

\bibitem{KS2009}
D. V. Khomitsky and E. Ya. Sherman,  Phys. Rev. B 79, 245321 (2009).

\bibitem{Rabi2012}
D.V. Khomitsky, L.V. Gulyaev, and E.Ya. Sherman,  Phys. Rev. B 85, 125312 (2012).

\bibitem{Chotorlishvili}
L. Chotorlishvili, Z. Toklikishvili, A. Komnik, et al,  Phys. Lett. A 377, 69 (2012);  J. Phys.: Condens. Matter 24, 255302 (2012).

\bibitem{Berggren}
K.-F. Berggren and T. Ouchterlony, Found. Phys. 31, 233 (2001).

\bibitem{Zinner2014}
O.V. Marchukov, A.G. Volosniev, D.F. Fedorov, et al,
J. Phys. B: At. Mol. Opt. Phys. 47, 195303 (2014).

\bibitem{billiard2013}
D.V. Khomitsky, A.I. Malyshev, E.Ya. Sherman, et al,  Phys. Rev. B 88, 195407 (2013).

\bibitem{BSSR}
K.-F. Berggren, A.F. Sadreev, and A.A. Starikov, Phys. Rev. E 66, 016218 (2002);
E.N. Bulgakov, D.N. Maksimov, and  A.F. Sadreev, Phys. Rev. E 71, 046205 (2005);
E.N. Bulgakov and I. Rotter, Phys. Rev. E 73, 066222 (2006);
K.-F. Berggren, D.N. Maksimov, A.F. Sadreev, et al, Phys. Rev. E 77, 066209 (2008).

\bibitem{Yang2007}
Z. Yang, S. Zhang, and Y. C. Li, Phys. Rev. Lett. 99, 134101 (2007).

\bibitem{BHZ}
B.A. Bernevig, T.L. Hughes, and S.-C. Zhang, Science 314, 1757 (2006);
M. K{\" o}nig, H. Buhmann, L.W. Molenkamp, et al, J. Phys. Soc. Jpn 77, 031007 (2008).

\bibitem{Timm}
C. Timm, Phys. Rev. B 86, 155456 (2012).

\bibitem{QDTI}
A. Kundu, A. Zazunov, A.L. Yeyati, et al, Phys. Rev. B 83, 125429  (2011);
G. Dolcetto, N. Traverso Ziani, M. Biggio, et al, Phys. Rev. B 87, 235423 (2013);
G.J. Ferreira and D. Loss, Phys. Rev. Lett. 111, 106802 (2013);
C. Ertler, M. Raith, and J. Fabian, Phys. Rev. B 89, 075432 (2014).

\bibitem{grapheneqd}
G. Giavaras, P.A. Maksym, and M. Roy, J. Phys. Cond. Mat. 21, 102201 (2009);
G. Giavaras and F. Nori, Phys. Rev. B 83, 165427 (2011); Phys. Rev. B 85, 165446 (2012).

\bibitem{MMbooks}
E.Y. Tsymbal and I. Z{\" u}tic (Editors), \textit{Handbook of Spin Transport and Magnetism}, CRC Press, Taylor and Francis Group, Boca Raton, 2012;
J.P. Liu, E. Fullerton, O. Gutfleisch, D.J. Sellmyer (Editors),
\textit{Nanoscale Magnetic Materials and Applications},
Springer Science+Business Media, New York, 2009;
{\' E}. du Tr{\' e}molet de Lacheisserie, D. Gignoux, M. Schlenker (Editors),
\textit{Magnetism. Materials and Applications}, Springer Science+Business Media, Boston, 2005.

\bibitem{Merkin}
D.R. Merkin, \textit{Introduction to the Theory of Stability}, Springer-Verlag New York, Inc., 1997.

\bibitem{Choi2014}
S. Choi, C.-H. Park, and S.G. Louie, Phys. Rev. Lett. {\bf 113}, 026802 (2014).

\end{thebibliography}
\end{document}